\documentclass[final,1p,11pt]{elsarticle}

\usepackage{epsfig}
\usepackage{array,tabularx,epsfig,mathrsfs,graphicx,rotating}
\usepackage{ifthen}
\usepackage{amsfonts}
\usepackage{ragged2e}
\PassOptionsToPackage{hyphens}{url}
\usepackage[hyphens]{url}
\usepackage{hyperref}
\usepackage{listings}
\usepackage{lineno}
\usepackage{subfig}
\usepackage{epstopdf}
\usepackage{color}
\usepackage{float}
\usepackage{verbatim}
\usepackage{color,soul}

\usepackage[normalem]{ulem} 
\usepackage{soul} 
\usepackage{amsmath,amssymb}

\let\originallesssim\lesssim
\let\originalgtrsim\gtrsim

\DeclareRobustCommand{\lesssim}{%
  \mathrel{\mathpalette\lowersim\originallesssim}%
}
\DeclareRobustCommand{\gtrsim}{%
  \mathrel{\mathpalette\lowersim\originalgtrsim}%
}

\makeatletter
\newcommand{\lowersim}[2]{%
  \sbox\z@{$#1<$}%
  \raisebox{-\dimexpr\height-\ht\z@}{$\m@th#1#2$}%
}
\makeatother

\hypersetup{
  colorlinks=true,
  linkcolor=blue,
  citecolor=blue,
  urlcolor=blue
}

\graphicspath{{figs/}}

\newcommand{\beq}{\begin{equation}}
\newcommand{\eeq}{\end{equation}}

\newcommand{\mjj}{m_{jj}}
\newcommand{\mll}{m_{ll}}
\newcommand{\mjjl}{m_{jjl}}
\newcommand{\mjjjl}{m_{jjjl}}
\newcommand{\mjjll}{m_{jjll}}

\newcommand{\fbb}{\mathrm{fb^{-1}}}
\newcommand{\abb}{\mathrm{ab^{-1}}}

\chardef\til=126

\journal{ANL-HEP-166648}

\begin{document}
\definecolor{mygreen}{rgb}{0,0.6,0} \definecolor{mygray}{rgb}{0.5,0.5,0.5} \definecolor{mymauve}{rgb}{0.58,0,0.82}

\lstset{ %
 backgroundcolor=\color{white},   
 basicstyle=\footnotesize,        
 breakatwhitespace=false,         
 breaklines=true,                 
 captionpos=b,                    
 commentstyle=\color{mygreen},    
 deletekeywords={...},            
 escapeinside={\%*}{*)},          
 extendedchars=true,              
 keepspaces=true,                 
 frame=tb,
 keywordstyle=\color{blue},       
 language=Python,                 
 otherkeywords={*,...},            
 rulecolor=\color{black},         
 showspaces=false,                
 showstringspaces=false,          
 showtabs=false,                  
 stepnumber=2,                    
 stringstyle=\color{mymauve},     
 tabsize=2,                        
 title=\lstname,                   
 numberstyle=\footnotesize,
 basicstyle=\small,
 basewidth={0.5em,0.5em}
}


\begin{frontmatter}

\title{
Model-independent searches for new physics in multi-body invariant masses}

\author[add1]{S.V.~Chekanov}
\ead{chekanov@anl.gov}

\author[add1]{S.~Darmora}
\ead{sdarmora@anl.gov}

\author[add3,add4]{W.~Islam}
\ead{wasikul@okstate.edu}

\author[add2,add1]{C.E.M.~Wagner}
\ead{cwagner@hep.anl.gov}

\author[add1]{J.~Zhang }
\ead{zhangjl@anl.gov}

\address[add1]{
HEP Division, Argonne National Laboratory,
 9700 S.~Cass Avenue,
Lemont, IL 60439, USA.
}

\address[add2]{
Physics Department, EFI and KICP, University of Chicago, Chicago, IL 60637, USA.
}

\address[add3]{
Department of Physics, Oklahoma State University, Stillwater, OK 74078, USA.
}

\address[add4]{
Department of Physics, University of Wisconsin, Madison, WI 53706, USA.
}

\begin{abstract}
Model-independent searches for physics beyond the Standard Model typically focus on invariant masses of two objects (jets, leptons
or photons). In this study we explore opportunities for similar model-agnostic searches in multi-body invariant masses.
In particular, we focus on the situations  
when new physics can be observed in a model-independent way in 
three- and four-body invariant masses of jets and leptons.
Such searches may have good prospects in finding new physics 
in the situations when  two-body invariant masses, that have been extensively explored at collider experiments in the past, 
cannot provide sufficient signatures for experimental observations.
\end{abstract}

\end{frontmatter}


\section{Introduction}

In spite of the great effort put into the search for  new physics at the Large Hadron Collider (LHC),
no signs of physics beyond the Standard Model (BSM)  have yet emerged.
If BSM physics exists within the reach of the LHC,  it is possible that its experimental
signatures are more complicated than originally anticipated. 
A popular method of finding new physics is
to study signal-like deviations in two-body (2-body) 
invariant masses of jets, leptons or photons, or studies of event rates, 
assuming that the Standard Model (SM) background rates are well understood in terms of
Monte Carlo (MC) simulations or data-driven control regions. However, it is possible that new physics is hidden in complex event signatures, beyond single- and two-particle  distributions for which MC simulations  may not be 
reliable and/or have significant uncertainties.  

In particle-collision experiments, model-independent searches for signal-like
deviations in invariant-masses distributions of jets, 
leptons or photons are typically performed by establishing  a background hypothesis representing our best knowledge of the SM.   
Such expectations can be obtained without MC simulations using data-driven ``control regions'', i.e. regions of actual data with similar kinematic features but without BSM events.
An alternative approach for the description of  shapes of SM background distribution is 
to perform a fit of the entire mass spectra with some analytic function
or using numeric smoothing techniques.
After the background hypothesis is established, searches for BSM physics  
are performed by looking for deviations above the established background shape.
Previous searches for heavy resonances in dijet mass distributions $(\mjj$) using the techniques described above have been performed by ATLAS~\cite{highmass,Aaboud:2017yvp} and CMS~\cite{2017520,Khachatryan:2016ecr,Sirunyan:2018pas}.

Model-independent searches using the above-mentioned technique 
applied to 2-body invariant masses are rather popular at the LHC.
Only a handful number of LHC studies~\cite{Chatrchyan:2012uxa,Sirunyan:2018duw} went  
beyond the two-object mass distributions. In particular, searches in multi-body invariant masses that involve different objects (jets, leptons, photons etc.), without relying on MC predictions,  are almost non-existent.
In this respect, LHC data have not been fully explored in multi-body invariant masses with the same precision as in published papers with 2-body masses.

In this paper we investigate three- and four-body (3- and 4-body) invariant masses 
for BSM processes. We will discuss the importance of such studies when direct observations of BSM signals 
in 2-body decays (i.e. dijets, di-leptons or di-photons) are difficult.
The latter can be due to their large width, or due to 
significant backgrounds from inclusive multi-jet QCD events. 
A typical requirement for the partial width ($\Gamma$) and the mass ($m$) for a heavy particle   
leading to signals that can be observed in invariant masses using the model-independent  approach is $\Gamma/m<0.2$.
Other possible  difficulties with 2-body invariant masses are large SM background, or small
masses $(\mjj<0.5$~TeV) which are difficult to study due to trigger pre-scales. \cite{highmass,Aaboud:2017yvp,2017520,Khachatryan:2016ecr,Sirunyan:2018pas}. 

Recently, a class of models predicting wide $Z'$ resonances
that cannot easily be found using the conventional experimental methods based on 2-body invariant masses has been discussed in Ref. \cite{Accomando:2019ahs,Li:2019pag}.
In this paper we argue that BSM events with similar  broad  
states coupled to SM fermions can be identified using 
three- and four-body decays.

\section{Final state radiation in BSM models}
\label{fsr}

Due to a low trigger efficiency for jets with  momenta 
below about 0.5~TeV, two-jet invariant masses are difficult to describe at the LHC using analytic functions or smoothing techniques. 
As the result of such difficulties, 
many experimental searches focus  on high-mass regions (typically, above  1 TeV). 
To mitigate the trigger problem for low-mass searches, 
one can require an associated object that can be used for event triggering. This object can be a photon \cite{Aaboud:2019zxd,Sirunyan:2019sgo} or a jet \cite{Sirunyan:2017nvi,Sirunyan:2017dnz,Sirunyan:2018ikr,Sirunyan:2019vxa} from initial-state radiations.

Particles, such as photons and vector-bosons from final-state radiation of quarks
in the decays such as $Z'\to q\bar{q}$ is a valuable option for searches for new physics 
in 2-body invariant masses. In such cases, additional objects from final-state radiation, 
such as photons or leptons (from $W/Z$) can be used for both event triggering and for calculations of  
multi-body invariant masses used in searches. The rate of such events is expected 
to be significantly lower than that in the inclusive dijet case. The main benefit of such events is in the smoothness of expected background shape for low masses, and the existence of a statistically independent event sample \footnote{This independent sample can be constructed after inverting the quality cuts on the electromagnetic  object reconstruction} to construct
control regions. 
Experimentally, the inclusion of additional electromagnetic objects (such as leptons $l$ or photons $\gamma$) 
to dijet invariant mass  should not significantly increase the 
resolution of the reconstructed invariant masses.  Therefore, the searches in the masses such as $m_{jj\gamma}$ or $\mjjl$ may have advantages for low invariant-mass studies and, at least, should be considered on an equal footing with the dijet studies.

\section{Cascade decays of BSM particles}
\label{monte}

\begin{figure}
\begin{center}
   \includegraphics[width=0.45\textwidth]{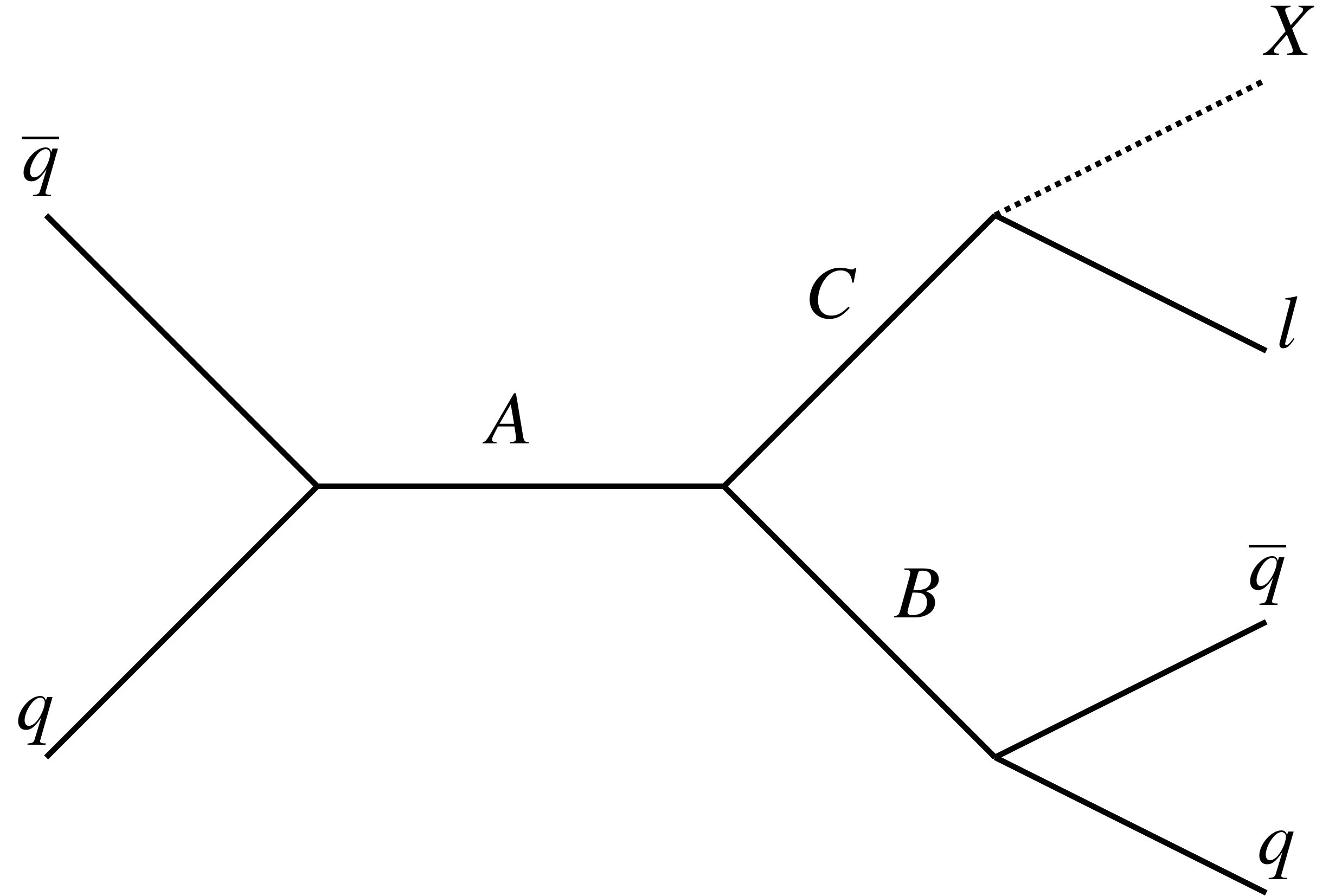}\hfill
\end{center}
\caption{A schematic representation of the decay a heavy particle $A$ to two other particles, $B$ and $C$.
}
\label{fig:1}
\end{figure}

Let us consider several event topologies which have a special property for model-independent 
BSM searches: such events might be  difficult to observe in 2-body decays, but many-body decays
can exhibit signals that can be detected through observations. 
To be more concrete, let us consider a cascade decay of a heavy particle $A$ into two other particles $B$ and $C$.
Such decays are typical for the diboson production (for a recent review see \cite{Dorigo:2018cbl}).
The most popular channel in experimental searches is when 
$B$ and $C$ are known bosons ($W$, $Z$ or $H$, on-shell or off-shell).
Generally, however, $A$ and $B$ may not be known.
In the case when $B$ and $C$ are bosons, their 2-body decay modes are ether hadronic ($q\bar{q}$) or leptonic ($l^{\pm}\nu$, $l^+l^-$). To be more specific, 
we will assume that $B$ decays to two jets, while $C$ decays
to a lepton and $X$, where $X$ is another lepton, neutrino or some other particle (detectable or undetectable).
This situation is depicted in Fig.~\ref{fig:1} where the dashed line from the particle $C$ represents such an additional particle.

Searches involving  jets and a lepton in the final state 
that originate from the cascade decays shown in Fig.~\ref{fig:1} was recently
studied \cite{Aad:2020kep} 
by ATLAS using a model-independent approach by triggering on leptons (electrons or muons), and by performing searches
in invariant masses,  $\mjj$, of two jets. A limitation for such an approach is the width of the heavy particle $B$: 
if $\Gamma/m >0.2$, the studies show a small sensitivity to such broad states since 
such deviations are almost indistinguishable  from the background hypothesis.

The studies \cite{Aad:2020kep} can be extended by allowing a lepton $l$  to 
participate in the reconstruction
of invariant masses. In this case, the three-body (3-body) invariant masses, $\mjjl$, 
can be calculated and studied
in exactly the same way as the dijet masses $\mjj$.  In  the situations when the original particle
$A$ has a small partial width,  the 3-body decays can exhibit a narrow signal, which can be sufficient for observation
of this class of BSM processes. Thus, even though particle $B$ is not directly observable in dijet decays due to its wide width, the original particle $A$ still can be found by observing
a deviation on a smooth background distribution
using a model-independent approach based on a global fit (or a numeric smoothing technique).
Such a class of physics studies has not been extensively explored at the LHC.
Below we will illustrate that the expected SM background for 3-body decays is indeed
a smoothly falling distribution.

\subsection{Monte Carlo simulations}
\label{mcpythia}

In this section we will illustrate several physics scenarios where 3- and 4-body invariant masses can be a useful tool for observing new physics. For our studies we will use the PYTHIA~8 MC model
\cite{Sjostrand:2006za, Sjostrand:2007gs} for generation of
$pp$-collision events at $\sqrt{s}=13$~TeV centre of mass energy. Effects from double parton scattering are not included in the simulation.
The NNPDF 2.3 LO \cite{Ball:2014uwa} parton density function
from the LHAPDF library \cite{Buckley:2014ana} was used.
Jets, isolated electrons and muons  were reconstructed from stable particles.
Jets were constructed with the anti-$k_T$ algorithm \cite{Cacciari:2008gp} as implemented in the {\sc FastJet} package~\cite{Cacciari:2011ma} using a distance parameter of $R=0.4$.
The minimum transverse energy of all jets was $40$~GeV in the pseudo-rapidity range of $|\eta|<2.5$.
The leptons are required to be isolated using 
a cone of the size $0.2$ in the azimuthal angle and pseudo-rapidity 
defined around the true direction of the lepton. All energies of particles inside this cone are summed.
A lepton is considered to be isolated if it carries more than $90\%$ of the cone energy.
All other details of simulation and reconstruction are described in \cite{Chekanov_2018}.

\subsection{Three-body invariant masses}

Let us discuss experimental situations where 3-body invariant masses can be more useful than the usual 2-body masses. 
Three-body invariant masses $\mjjl$ reconstructed from two jets and a lepton can be a more powerful 
variable in the identification of the process $A\rightarrow B\, C$ than 2-body masses under the following conditions:

\begin{itemize}

\item Two  jets from decays  of the particle $B$ should be resolved using the standard jet algorithms. 
See Fig.~\ref{fig:1}.
The angular separation of two jets can be approximated by $2 m_B/p_T^B$.                 
This implies that the mass
of $B$ should be comparable with the mass of $A$, such that the particle $B$ will not 
receive a significant boost in transverse momentum, thus the two jets from the particle $B$ can be well resolved.

\item 
Presence of at least one lepton in an event does
not set any particular constraints on the mass of $C$ compared to the mass of $A$.
For example, the decay of particle $C$ can be considered fully boosted, i.e.
two leptons originating from $C$ will be merged into a single lepton as seen by experiments. 
This can happen when the particle $C$ has a significantly
smaller mass compared to $A$, thus the particle $C$ receives a significant boost in momentum. For a TeV scale searches of the particle $A$, the SM $Z$ boson with the fully boosted decay to $l^+l^-$ can be a 
possible candidate for $C$.

There is another option for this topology: $C \rightarrow l+X$, where $X$ is an undetectable particle, as shown in Fig.~\ref{fig:1}. The particle $C$ can be a $W$ boson, or a BSM boson-like particle, while $X$ can be a neutrino. Such cases also do not impose any restrictions  on the mass of $C$ relative to the mass of $A$.

\end{itemize}  
 
The  above scenarios are convenient for model-independent searches
using a  data-defined  (or phenomenological) smoothly-falling shape for the $\mjjl$ distribution 
if the partial width of $A$ is narrow,
i.e. $\Gamma/m<0.2$.   
The advantage of searches in $\mjjl$, compared to the direct observation of $B$ in the dijet masses $\mjj$, will
be most obvious in the case when the partial width of $B$ is large (as in the case of strong decays).

Another possible combination for 3-body invariant masses is $j+ll$.
If a particle $C$ decays to two resolved leptons then this type of decays should be detected via the direct searches
in di-lepton channels. Thus this channel should be well covered by the existing LHC studies.
On the other hand, if $C$ decays to a highly boosted di-lepton pair which is resolved as a single lepton,
and particle $B$ decays to a jet and  lepton, this study can be partially covered by
the searches for quantum black holes  \cite{Aad:2013gma} that use the $m_{jl}$ invariant masses.

\section{Specific BSM models}

In this section we will consider several specific BSM models that can be found using 3-body invariant masses.

\subsection{Models with SM bosons}

\begin{figure}[h]
  \begin{center}
    \includegraphics[width=0.45\textwidth]{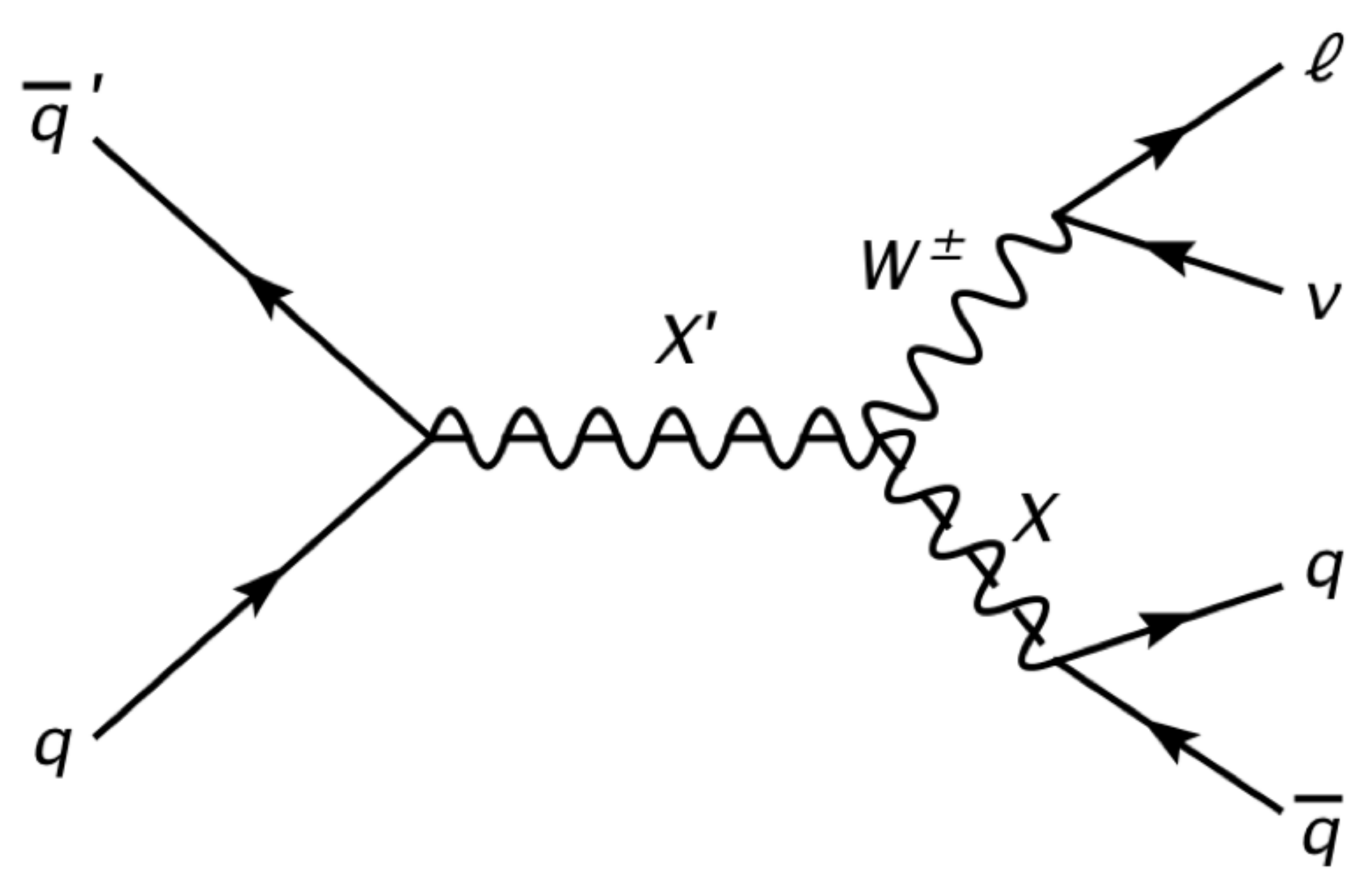}
  \end{center}
  \caption{A representative Feynman diagram for a generic resonance, 
  $X$, decaying to two partons in
  association with a leptonically decaying a $W$ boson in the $s$-channel.
  The latter channel includes a resonance $X'$  decaying to a $W$ and an $X$.}
  \label{fig:feynman}
\end{figure}

A cascade decay of two heavy particles with the associated $W$ gauge boson decaying to $l\nu$ is the natural class of events where multi-body invariant masses are useful for BSM searches.
A generic diagram of such processes is shown in Fig.~\ref{fig:feynman}.

As an example, let us consider  the processes $W'\to W Z' \to \ell\nu q\bar{q}$ and $\rho_{T} \rightarrow W^{\pm} \pi_{T} \to \ell\nu q\bar{q}$ process discussed in \cite{Aad:2020kep}.
The $W'\to W Z' \to \ell\nu q\bar{q}$ and $\rho_{T} \rightarrow W^{\pm} \pi_{T} \to \ell\nu q\bar{q}$
simulations assume a $Z'$ dijet resonance produced in association with
a leptonically decaying $W$ from the $q\bar{q}\to W'$ process. The mass difference between the
$W'$ and $Z'$ was set to 250~GeV. The latter requirement yields the largest predicted cross-section
for the desired final state \cite{Aad:2020kep}.

The second process discussed in \cite{Aad:2020kep} is $\rho_{T} \rightarrow W^{\pm} \pi_{T} \to \ell\nu q\bar{q}$. It is a generic technicolor model with a technirho, $\rho_{T}$, that decays into a leptonically decaying $W$ boson and a technipion $\pi_{T}$, decaying into two jets.
The mass of the $ \rho_{T}$ is chosen to be larger than the mass of the $\pi_{T}$ by a factor of two,
which maximizes the cross-section for the $l\nu q\bar{q}$ final state.
These processes are quite distinct kinetically: in one case the phase space for $W$ production is limited, while it  increases with the mass of $\rho_{T}$ for the technicolor model.

\begin{figure}
\begin{center}
  
   \subfloat[$W'/Z'$ SSM model] {
   \includegraphics[width=0.55\textwidth]{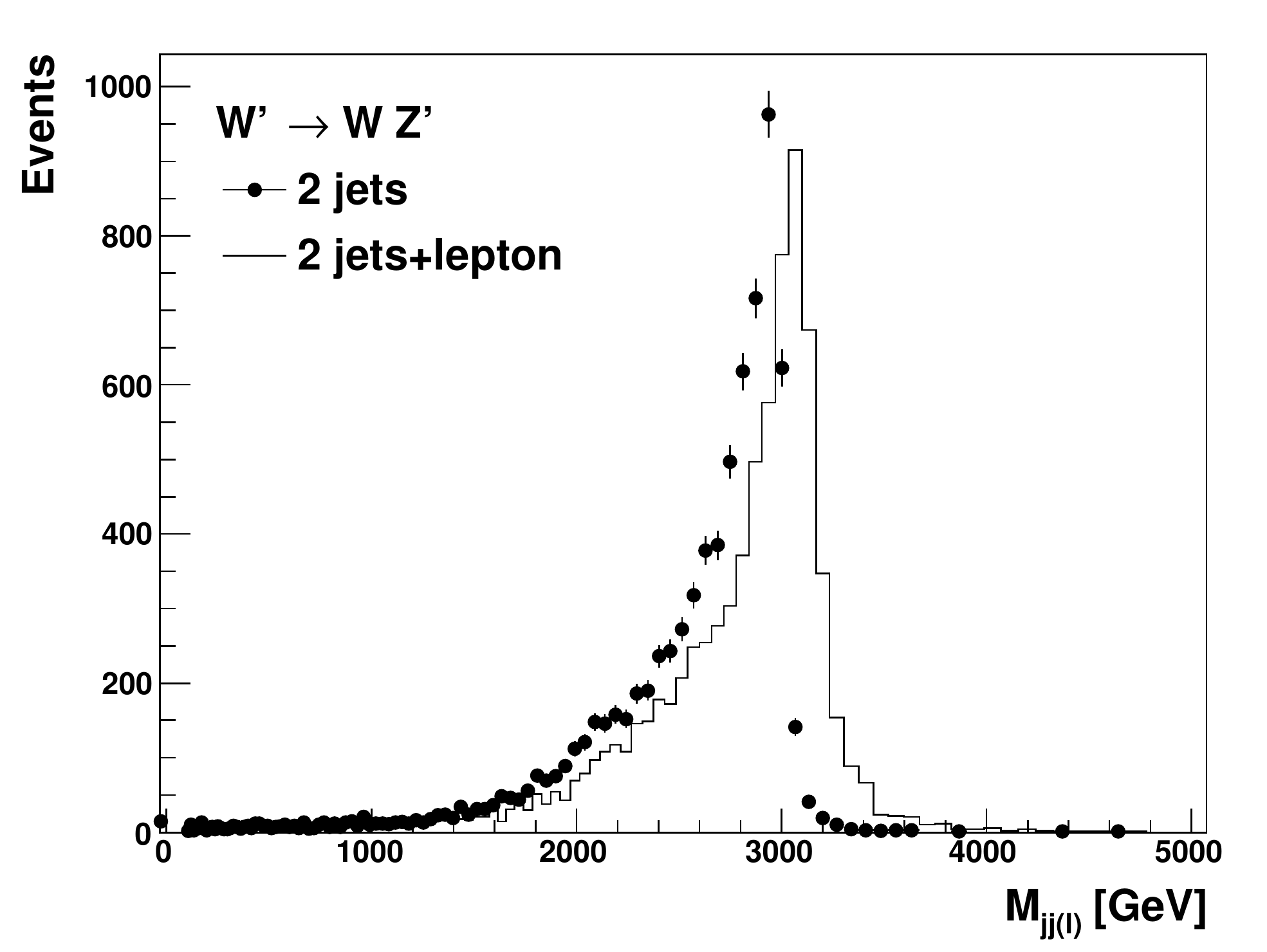}
   }
   
   \subfloat[$\rho_{T}$ model] {
   \includegraphics[width=0.55\textwidth]{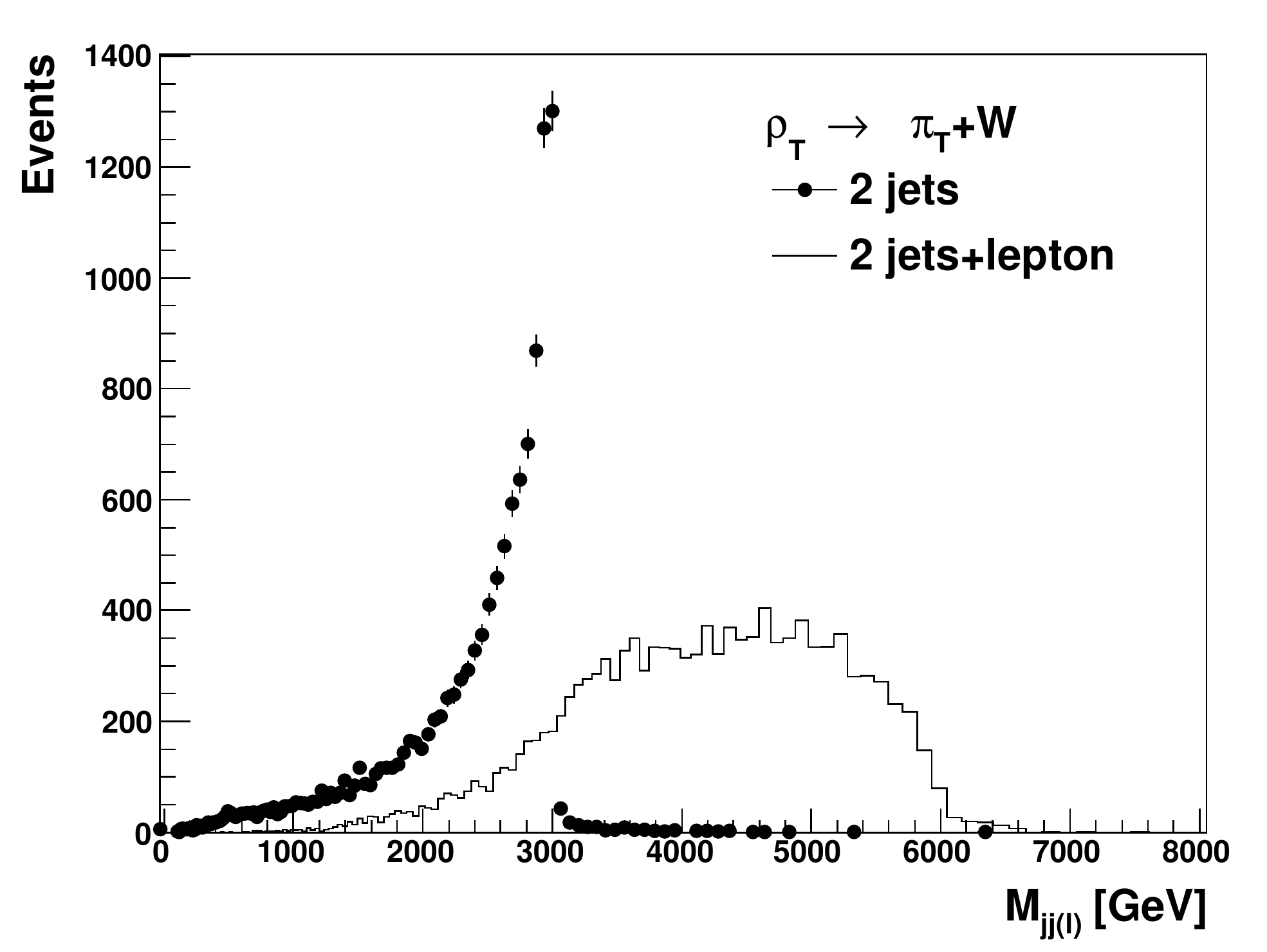}
   }

\end{center}
\caption{The invariant masses of two jets ($\mjj$) and two jets and a lepton ($\mjjl$) 
for (a)  $W'\to W Z' \to \ell\nu q\bar{q}$ and (b) $\rho_{T} \rightarrow W^{\pm} \pi_{T} \to \ell\nu q\bar{q}$ process discussed in \cite{Aad:2020kep}. The SSM  process has  a mass splitting between $W'$ and $Z'$ of 250~GeV. The second model has a mass of the $\rho_T$ twice larger than that of the $\pi_T$. 
}
\label{fig:wz}
\end{figure}

Figure~\ref{fig:wz} shows the 2-jet and 2-jet+lepton invariant masses for the processes
discussed above. The processes were simulated and reconstructed as discussed in Sect.~\ref{monte}.
The result shows that there is no principal difference in the width of the 2-jet and 3-body invariant mass
for  $W'\to W Z' \to \ell\nu q\bar{q}$. 
The relative  widths of these masses are below 11\% (it was estimated by taking the 90\% Root-Mean Squared value (RMS) of the distribution and dividing by the mass).  Thus, the  effect from the missing neutrino is small.
This is not the case for 
$\rho_{T} \rightarrow W^{\pm} \pi_{T} \to \ell\nu q\bar{q}$ where the width of the 3-body mass is significantly larger than that for the 2-jet mass. 

These examples show that, in some scenarios, like the first one discussed above, the discovery potential using 3-body masses may be as strong as when using the two-jet masses. Such scenarios belong to the cases when the phase space  for production of additional particles, such as $W$, is restricted, and contribution from missing neutrino can be approximately ignored. In the unrestricted case, instead,
the inclusion of missing energy (MET) from escaping neutrinos leads to broad mass 
distributions, thus  preventing an effective search in smoothly decreasing  distributions. 
Such broad distributions are due to the fact that the reconstructed transverse mass 
distribution is, in general, much wider than the invariant mass.

\subsection{Radion models}

Generally, in order to identify a scenario where 3-body masses have a strong potential for narrow signal signatures,  a scan of free 
parameters needs to be  performed.  The main tunable parameter is the mass splitting between 
the two heavy BSM particles. As an example
of this scenario, let us consider the following model \cite{Agashe:2016kfr}:
$$
W_{kk} \to W + \varphi \to l\nu + gg  
$$
where $W_{kk}$ denotes the Kaluza-Klein excitation  ("KK boson"), 
and $\varphi$ is a radion particle decaying to two gluons (i.e jets).
Since the $W$ bosons decay into $l\nu$, the events should include a $\nu$ leading to missing transverse energy (MET). 
Adding reconstructed MET to the invariant mass reconstruction significantly increases the observed signal width.
Therefore, as in the first example in the last section, we shall explore physics scenarios when MET is not required 
(i.e. when the transverse energy of $\nu$ is not large).

\begin{figure}
\begin{center}

\subfloat[Case 1] {
\includegraphics[width=0.45\textwidth]{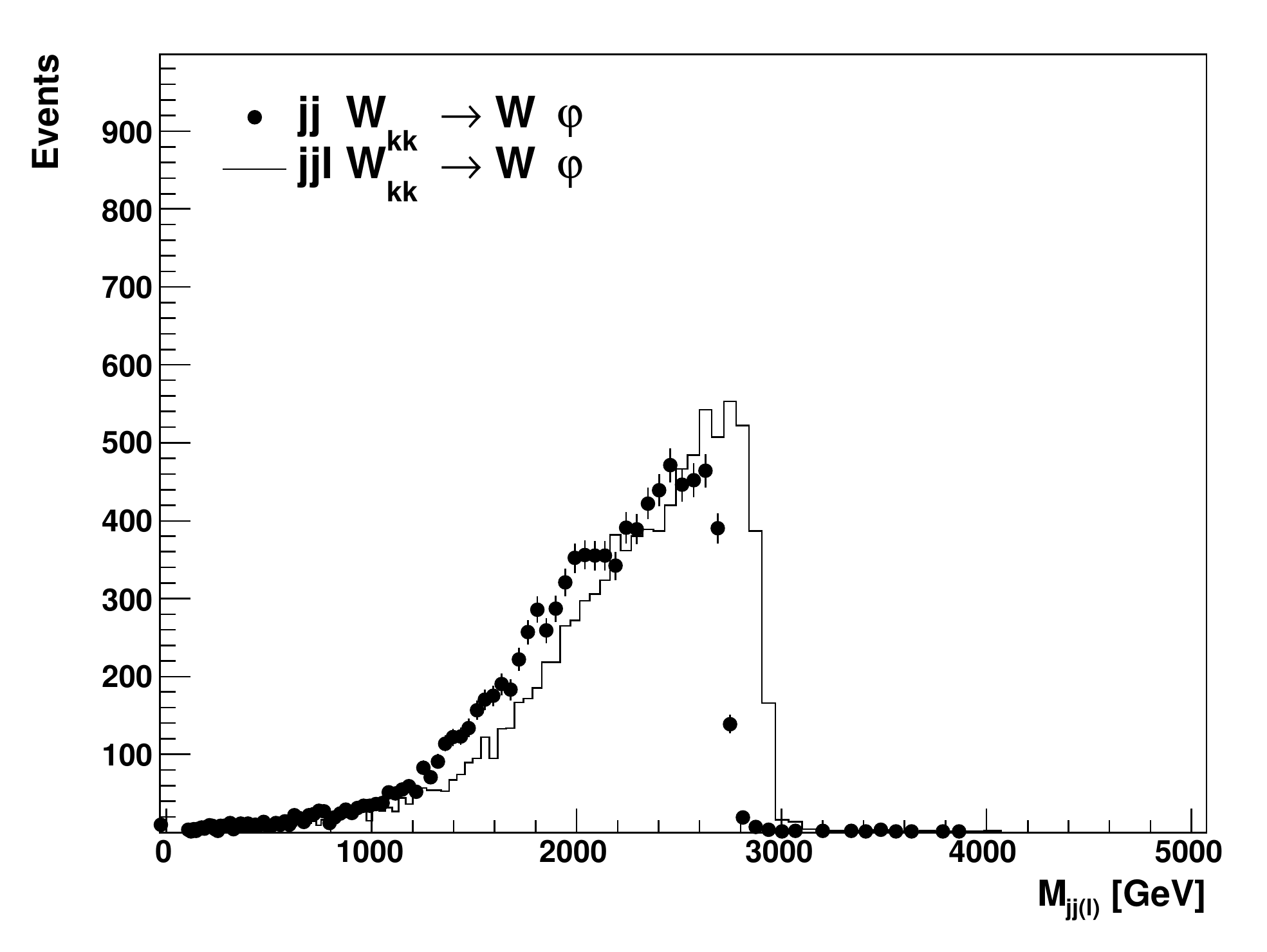}\hfill               
}

\subfloat[Case 2] {
\includegraphics[width=0.45\textwidth]{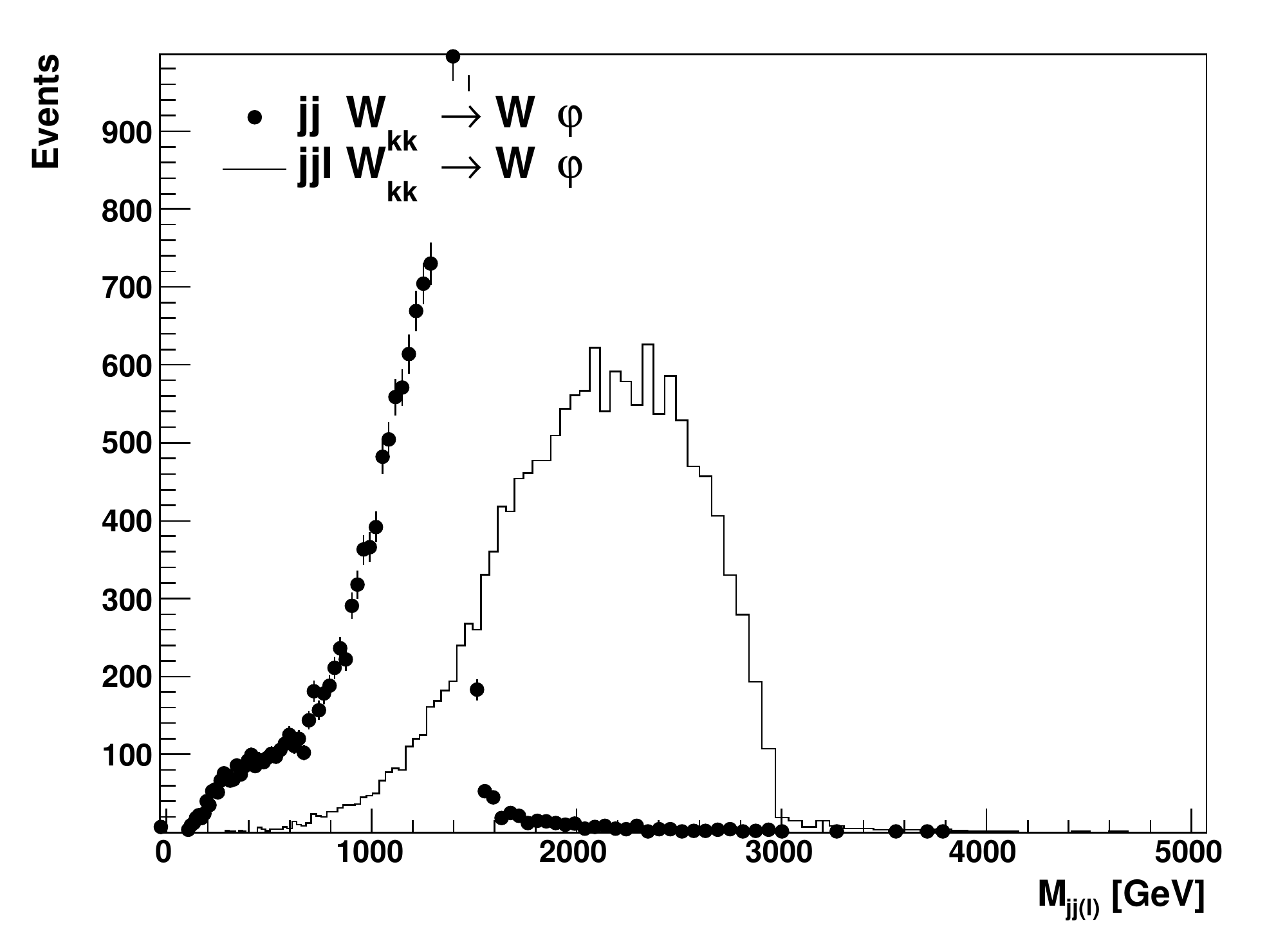}              
}

\end{center}
\caption{The $\mjj$  and $\mjjl$  invariant mass distributions for the 3 TeV $Wkk$ mass when (a) case 1 : $M_{W_{kk}} - M_{\varphi} = 250$ GeV \& (b).
Case 2 : $M_{\varphi} = M_{W_{kk}} / 2$ . }
\label{fig:radion_case_comparison}
\end{figure}

In this study we varied the masses of $W_{kk}$ and $\varphi$, and found a case when the width of 3-body
invariant mass is the same as in the case of 2-body decays (that corresponds
to two-jets from the radion).
Figure~\ref{fig:radion_case_comparison}(a) shows the scenario where 
the masses difference between $W_{kk}$ and $\varphi$ is 250~GeV, i.e.  $M_{W_{kk}} - M_{\varphi} = 250$ GeV.
This leads to the width of $\mjjl$ close to the with of $\mjj$.

Another, alternative scenario, is shown in Figure~\ref{fig:radion_case_comparison}(b)
where $M_{\varphi} = M_{W_{kk}} / 2$. In the latter case, MET should not be ignored since 
the neutrino has a significant transverse momentum. It should be noted that, at large $W_{kk}$  masses,
the cross section for the second scenario shown in Figure~\ref{fig:radion_case_comparison}(b)
is a factor 2-5 larger than for the first scenario. However, the benefit of having a narrow signal width for $\mjjl$, leading to large reconstruction sensitivity over a smooth SM background, can compensate for effects of a lower absolute cross section. 
Therefore, the scenario with mass splitting between the two BSM particles close to the mass of associated particles (such as $W/Z/H$) is a preferable configuration for experimental searches in 3-body decays.

\subsection{Composite resonances}

\begin{figure}[h]
  \begin{center}
    \includegraphics[width=0.5\textwidth]{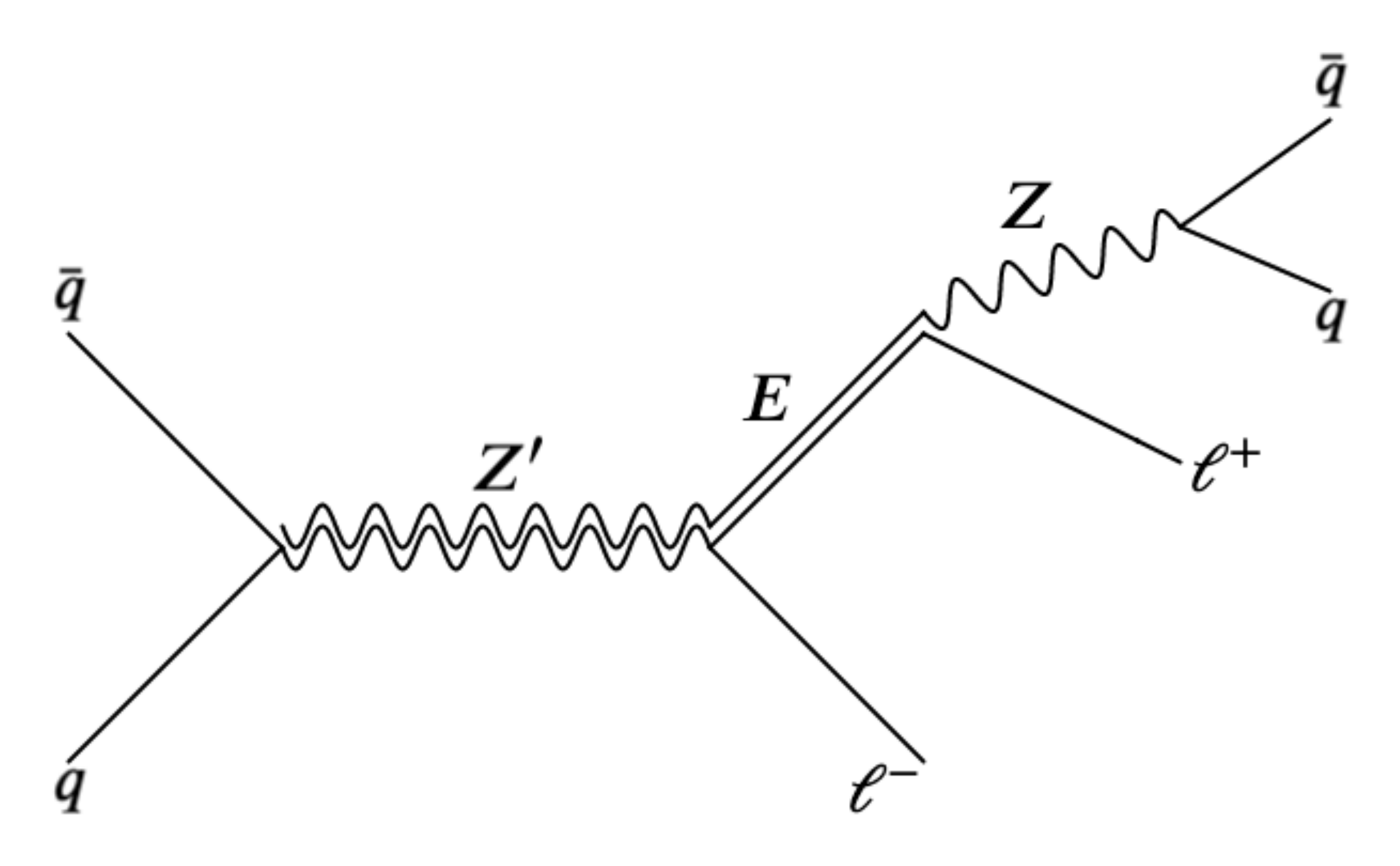}
    \label{fig:feynman_schan}
  \end{center}
  \caption{A representative Feynman diagram for the composite lepton model (see Sect. IV of \cite{Chala_2018}) where $\mjjl$ and $\mjjll$ can 
be used for searches of $V=Z'$ and a composite lepton-like particle $E$.}
  \label{fig:compos}
\end{figure}

Among different composite resonances models that break lepton flavour (LF) universality,
the decays channel $pp\to V \to E^{\pm} l^{\mp}$ (where $E$ is a heavy composite lepton which decays into SM bosons and leptons $E\to Z/h + l$) \cite{Chala_2018} 
can easily be probed using the multi-body invariant masses.
Figure~\ref{fig:compos} shows a Feynman diagram for a model where $V$ is a heavy $Z'$ boson ($M(Z')>M(E)$).
When $Z/h$ decays hadronically, invariant masses of 2 jets and a lepton ($\mjjl$) can be used to reconstruct the mass of $E$.
In addition, the 2j+2l mass ($\mjjll$) is sensitive  to the $V$ ($Z'$)  mass itself. It should be noted that even if the partial width of the composite $E$ is broad (and thus
$\mjjl$ is not sensitive to the presence of $E$), the invariant mass $\mjjll$ can still be narrow if the partial width of $V$ is small (which is a typical case for $Z'$).

\begin{figure}[h]
  \begin{center}
    \includegraphics[width=0.7\textwidth]{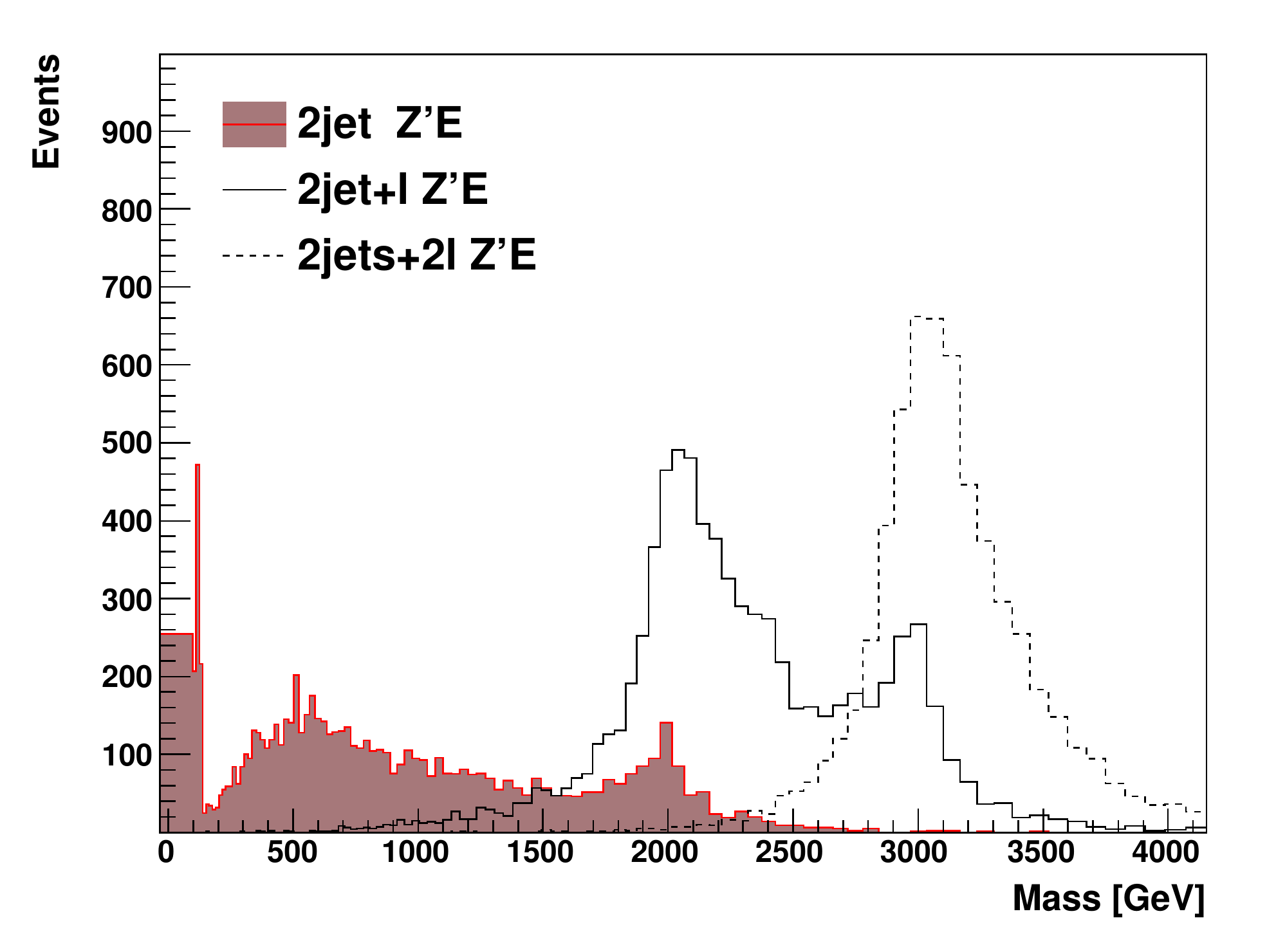}
    \label{fig:feynman_schan}
  \end{center}
  \caption{Invariant masses for a $Z' E$ model \cite{Chala_2018} constructed from 2 jets, 2 jets plus a lepton and 2 jets + 2 leptons. The $Z'$ mass was generated at 3 TeV, while $M(E)=2$~TeV. The simulations are performed using MG5 with PYTHIA~8 showering.}.
  \label{fig:mass_compH}
\end{figure}

Figure~\ref{fig:mass_compH} shows the $\mjj$, $\mjjl$ and $\mjjll$ invariant masses
reconstructed at the truth level of a Monte Carlo simulation.
The selection cuts are set as for the default analysis used in this study. 
The simulations are performed using Madgraph5 (MG5) \cite{Alwall_2014} with the PYTHIA~8 shower.
The masses are set to  $M(Z')=3$~TeV and $M(E)=2$~TeV in the MG5 simulation.
The observed peak near  3~TeV (shown with dotted line) 
is due to the decay of $Z'$.
The solid line shows two peaks. The largest peak is a reflection of the
$E$ decays, while the smaller peak is a reflection of the $Z'$ decays.
The 2-jet masses show the mass peak at the $Z$ mass, but it has a more complicated distribution than multi-body decays.
This figure illustrates that both 2j+l and 2j+2l invariant masses can be used for searches
of events originating from the composite resonances models \cite{Chala_2018}.

\subsection{Hypothetical scenarios}

Based on our analyses in the previous sections, 
let us consider  hypothetical examples where 2-body invariant masses are at a disadvantage for searches since the 2-body signal width is larger than 3 or 4-body masses. The goal of such simulation is to illustrate a possible parameter space when 2-body masses
are less powerful for searching enhancements on a smoothly falling background than for multi-body masses.

\subsubsection{Three-body invariant masses}
\begin{figure}
\begin{center}
   \includegraphics[width=0.7\textwidth]{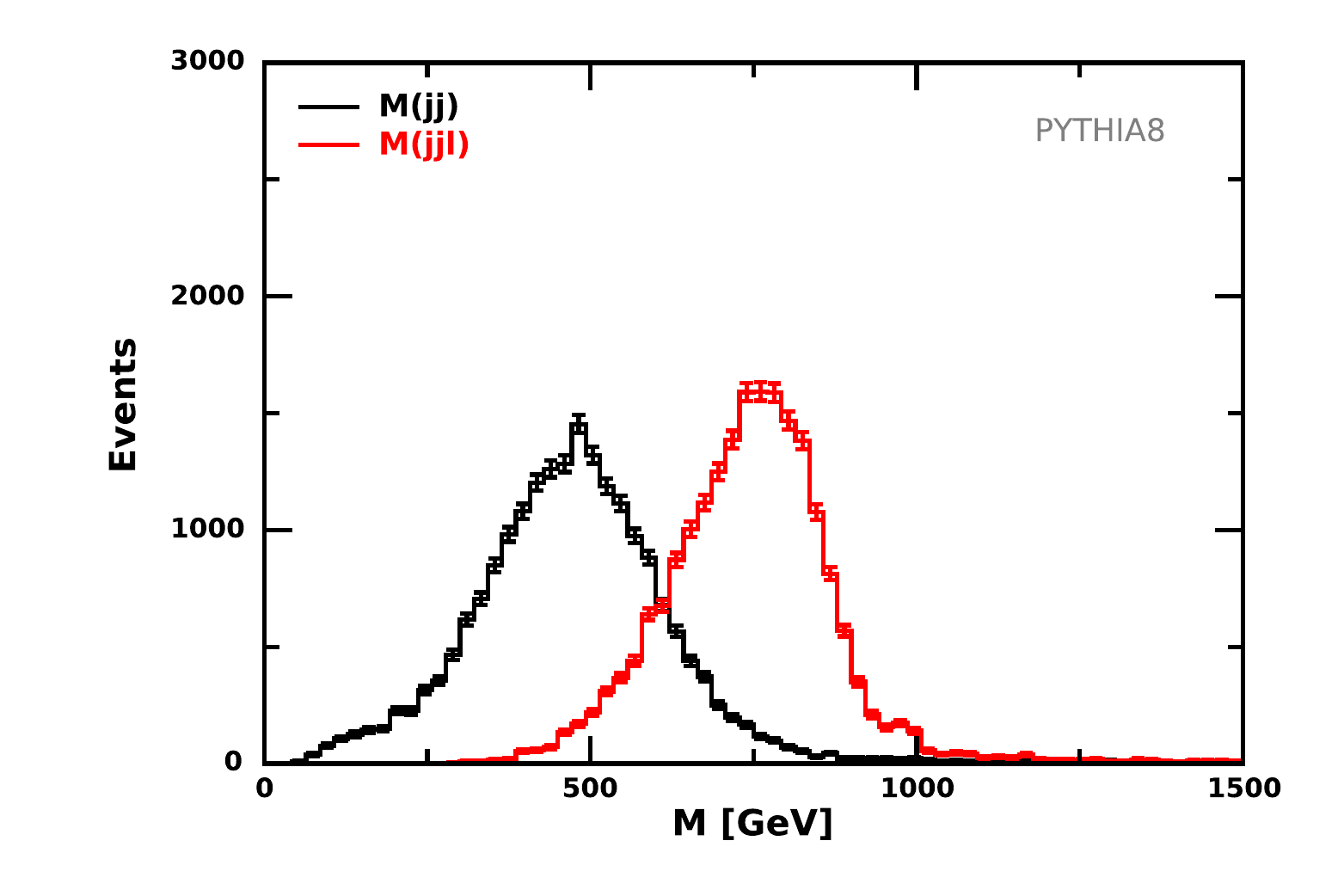}\hfill
\end{center}
\caption{The invariant masses of two jets ($\mjj$) and two jets and a lepton ($\mjjl$) in events $A\rightarrow B\, C$ simulated 
with PYTHIA~8 with the settings  described in the text. 
}
\label{fig:2}
\end{figure}

For the 3-body example, we  setup a PYTHIA~8 simulations for the process 
$W'\rightarrow Z'\, W^*$, where $W'$, $Z'$ and $W^*$ are hypothetical bosons, without assuming any particular physics couplings or model. 
In this simulation, the  mass of $W'$ was set to 1~TeV and its width to $\Gamma=10$~GeV.
$Z'$ decays to two jets, while $W^*$ decays to a lepton and a neutrino.
The mass of $Z'$ was set to 500~GeV (with $\Gamma=250$~GeV), and the mass
of $W^*$ was set to be of 300~GeV (with $\Gamma=250$~GeV).
The widths of $Z'$ and $W^*$ in this setup are large ($\Gamma/m\geq 0.5$)  and detection of such particles via the
observations of Gaussian-shaped enhancements on smoothly falling $\mjj$ distributions is  difficult.
However, for the $\mjjl$ masses, the situation is different.  Figure~\ref{fig:2} shows that the width of $\mjjl$ is smaller than the width of $\mjj$ despite the presence of the neutrino that introduces a shift and smearing in the mass distribution.  
According to this simulation, the relative width of the $\mjjl$ masses is close to  $15\%$,  which is in the range  of acceptable width for the experimental detection on smoothly falling background distributions.

\subsubsection{Four-body invariant masses}

For the 4-body decay analysis, let's consider the example shown in Fig.~\ref{fig:1}, where X could be a lepton or a jet. Hence, events with at least two jets and leptons lead to the following combinations: $\mjjll$ and $\mjjjl$.
As in the case of 3-body decays, model-independent searches for signal-like enhancements in
smoothly falling distributions  of 4-body invariant masses  can be simpler than for similar searches
in 2-body decays when the partial widths of $B$ and $C$ have $\Gamma/m>0.2$.

Here are possible kinematic scenarios that are not well covered in $pp$ collider experiments:

\begin{itemize}

\item
Signature searches based on $\mjjll$ invariant masses for discovering a TeV-scale particle decaying to two heavy particles when both $B$ and $C$ are heavy, thus there is no significant momentum boost, and all decay products are well resolved.

\item
Event topologies  contributing to $\mjjjl$, when a particle $C$ decays to a lepton and jets, similar
to quantum black holes scenarios \cite{Aad:2013gma}.      

\end{itemize}

\begin{figure}
\begin{center}
   \includegraphics[width=0.7\textwidth]{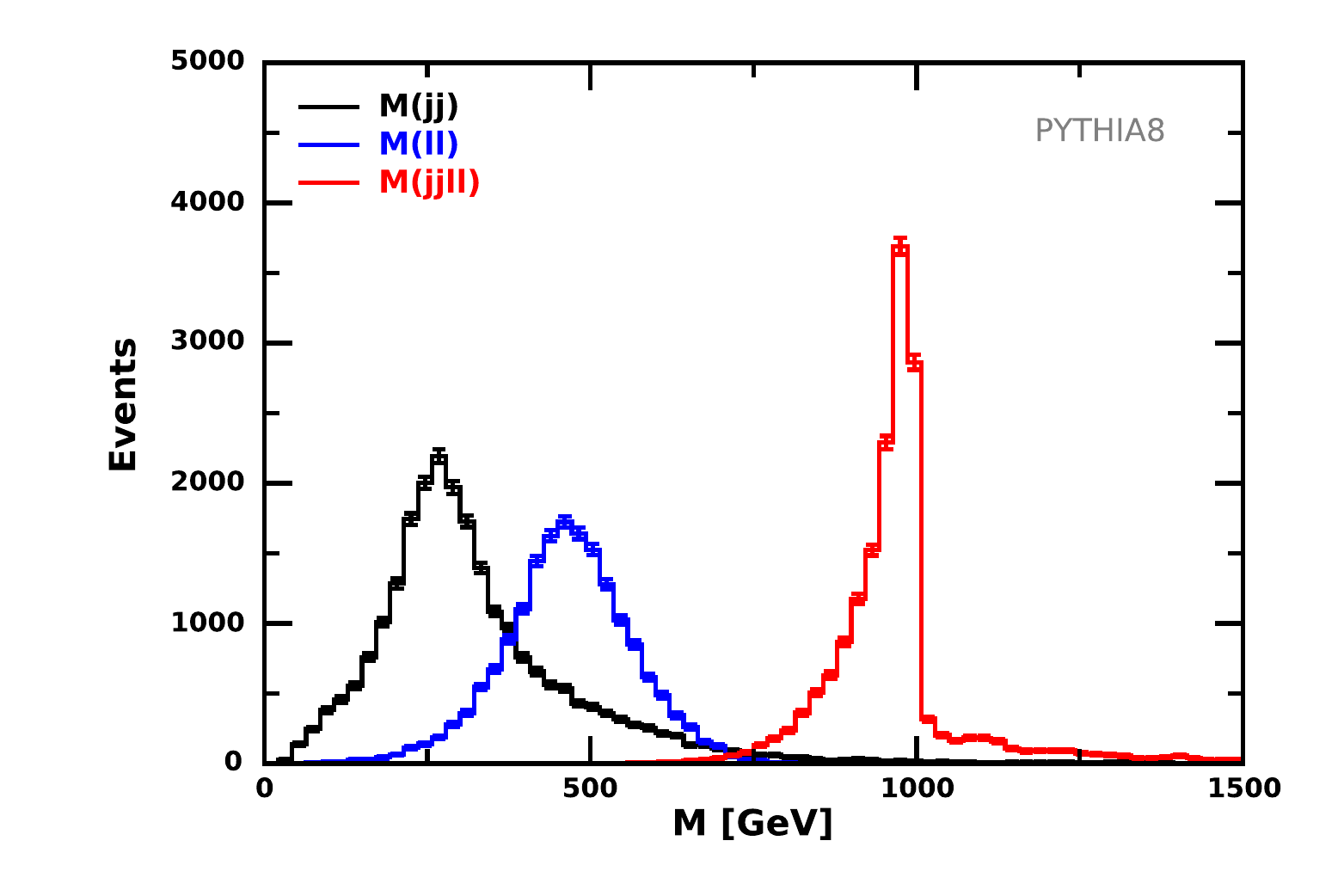}\hfill
\end{center}
\caption{The invariant masses of two jets ($\mjj$), two leptons ($\mll$)  and two jets and two leptons ($\mjjll$)
in events $A\rightarrow B\, C$ simulated  
by PYTHIA~8 with the settings as described in the text.
}
\label{fig:3}
\end{figure}

Let us consider a MC simulation for the first scenario.  We have created the PYTHIA~8 simulation
for the process $W'\rightarrow Z'\, W^*$, where $W'$, $Z'$ and $W^*$ are hypothetical bosons.
The mass of $W'$ was set to 1~TeV and the width $\Gamma=10$~GeV.
$Z'$ decays to two leptons (electrons or muons), while $W^*$ to two jets.  
The mass of $Z'$ was set to 500~GeV (with the width of 250~GeV), and the mass
of $W^*$ was set to 300~GeV (with the width of 150~GeV).
Thus the widths of $Z'$ and $W^*$ are broad,  and a detection of such particles via the
observations of Gaussian-shaped enhancements in 2-body decays is  difficult.
Figure~\ref{fig:3} shows that the relative width of $\mjjll$ is smaller than the widths of 2-body decays of $Z'$ and $W^*$.
Thus the relative width of the $\mjjll$ distribution due to the $W'$ decays is close to  the range
for a direct observation of such a state over a smoothly falling background.

\section{Exclusion limits}
Although it would be interesting to derive
the limits from  4-body invariant masses, such as $\mjjll$ or $\mjjjl$, obtaining these limits is difficult due to the lack of experimental publications that can be used
to correct Monte Carlo predictions for  realistic lepton misidentification rates.

To set upper limits on observation of a new resonance in the $\mjjl$ invariant mass,
Monte Carlo simulations have been performed using PYTHIA~8 with settings and object reconstruction as 
described in Section~\ref{mcpythia}.
Three SM processes were generated: (1) light-flavor QCD dijets, (2) vector and scalar boson production and (3) $t\bar{t}$.
Pileup events were not simulated, but this problem is mitigated by using the normalization factors described below.
All other details in the simulation match the description of \cite{Chekanov_2018}. Unlike the signal models,
the SM background processes require simulations of misidentification rates for muons and leptons (``fake rates``).
We use 0.1\% misidentification rate for muons, and 1\% fake rate for electrons \cite{Chekanov_2018}.
This is implemented by assigning the probability of $10^{-3}$ ($10^{-2}$)
for a jet to be identified as a muon (electron) using a random number generator.
The distributions were obtained for events having at least one isolated
lepton with $p_T^{l}>60$~GeV and two jets  at $p_T>30$~GeV.
The SM background MC samples are available from the HepSim repository \cite{Chekanov:2017pnx}. 

After the $\mjjl$ distributions were reconstructed  assuming 140~fb$^{-1}$ separately 
for muons and electrons, they were  scaled to match the official ATLAS
Monte Carlo\footnote{Such distributions are available as additional material from the ATLAS web page of the paper \cite{Aad:2020kep}}. This corrects for residual differences between the truth-level simulations
as described above and the ATLAS simulations that include realistic detector effects,  i.e. $\mjjl$-dependent 
lepton misidentification rates and pile-up contributions.  The values of the correction are 2 -- 3, depending  on $\mjjl$. These values are almost entirely attributed to the correction for the lepton 
misidentification rates.

\begin{figure}[h]
\begin{center}
  
   \subfloat[ $140~\fbb$] {
   \includegraphics[width=0.5\textwidth]{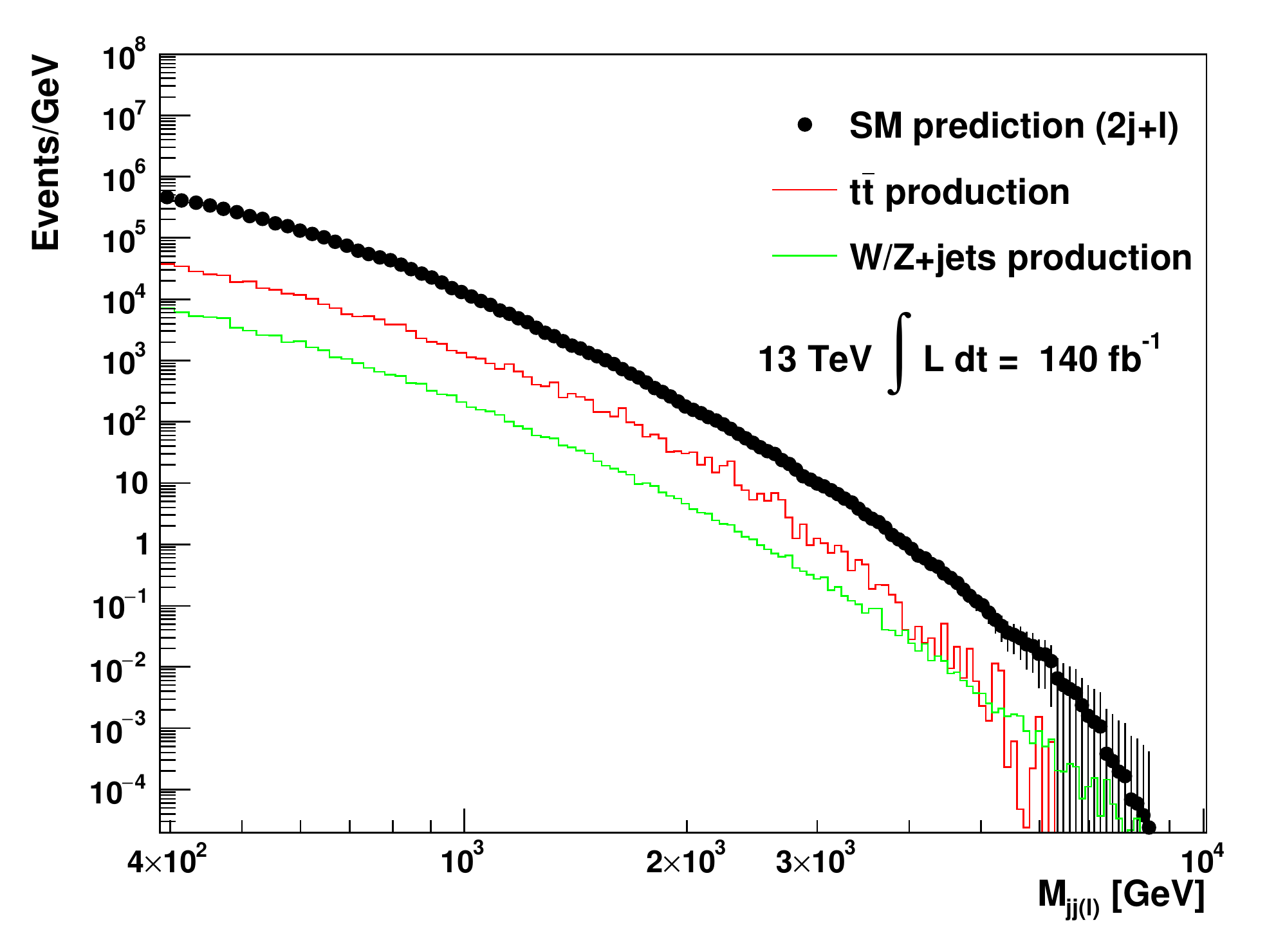}\hfill
   }
   
   \subfloat[ $440~\fbb$] {
   \includegraphics[width=0.5\textwidth]{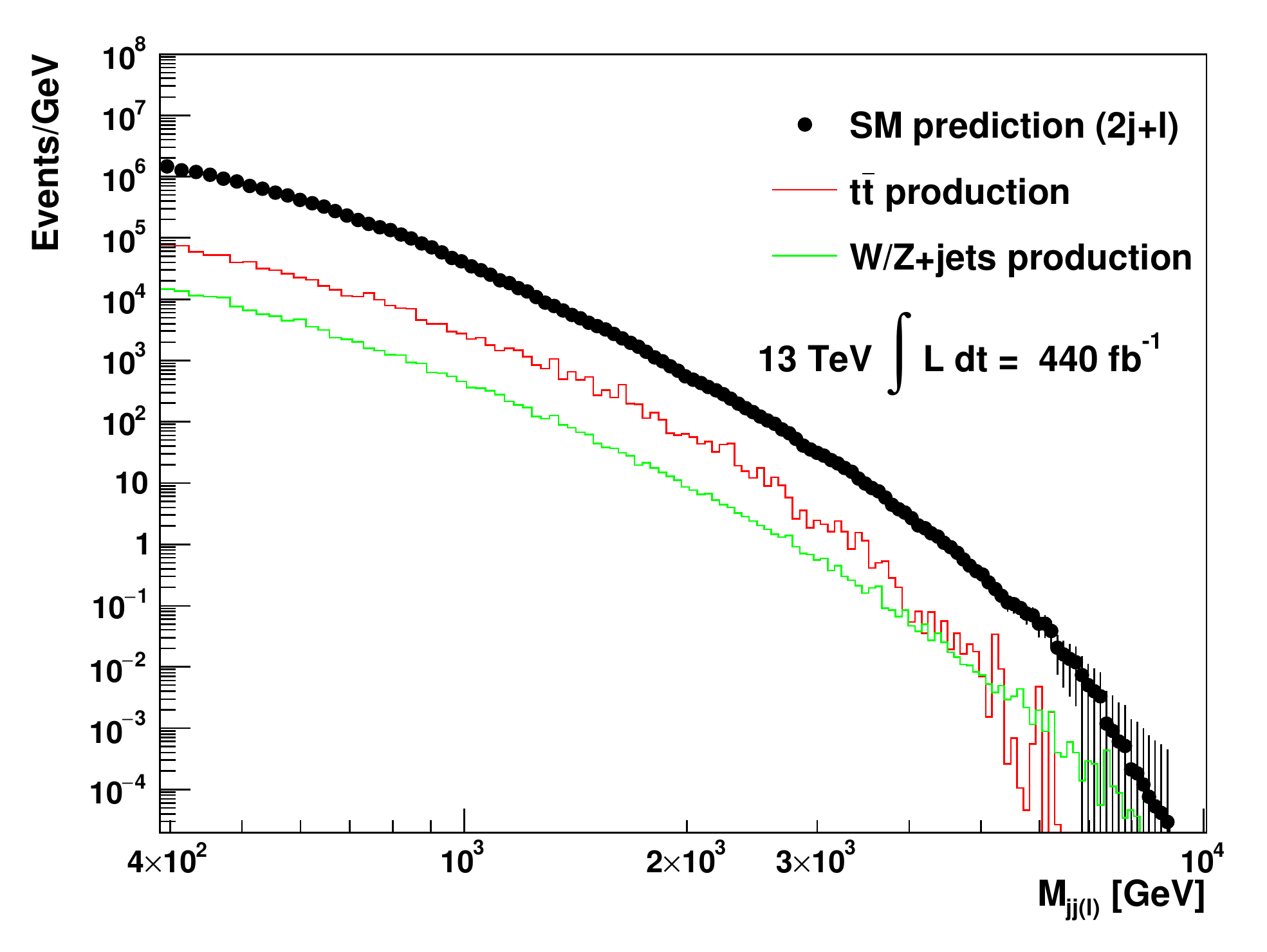}\hfill
   }
   
   \subfloat[ $3~\abb$] {
   \includegraphics[width=0.5\textwidth]{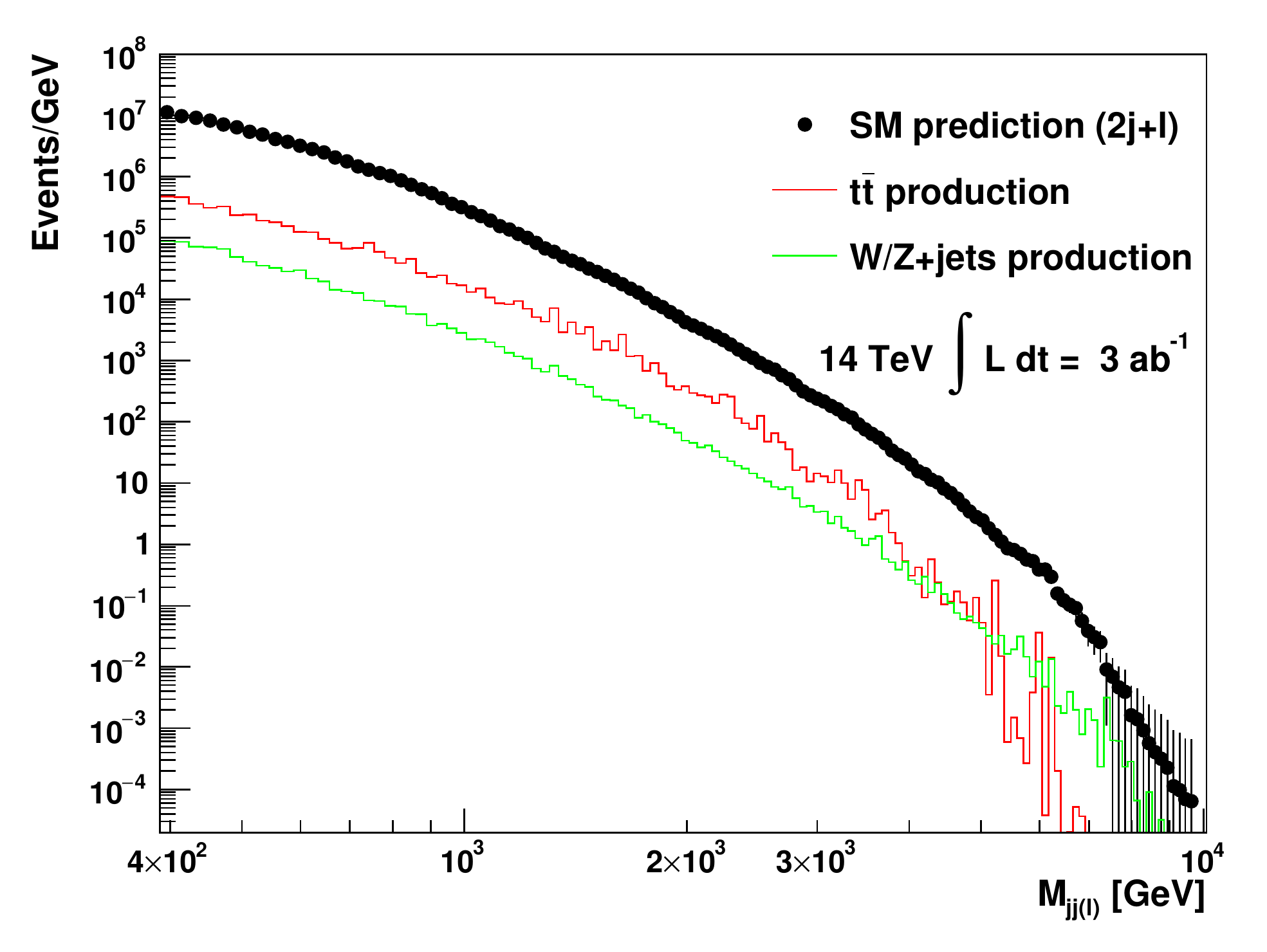}
   }
\end{center}
\caption{Expectations for $\mjjl$ invariant mass distributions for 140~$\fbb$, 440~$\fbb$ and  3~$\abb$ (14~TeV)
using the PYTHIA~8 generator for events having at least one isolated
lepton with $p_T^{l}>60$~GeV. Contributions
from  $W/Z/H^0$ -boson processes and top-quark processes are shown separately (without stacking the histograms). }
\label{fig:LHCmjjl}
\end{figure}

Figure~\ref{fig:LHCmjjl} shows the invariant masses using three luminosity scenarios: 
140~fb$^{-1}$, 440~fb$^{-1}$ (LHC Run 2+ Run 3) and the HL-LHC (3 ab$^{-1}$ for 14~TeV). 
The latter distribution was corrected by a correction factor of 12\% (estimated using PYTHIA~8) 
to reflect the change in the CM energy from 13 to 14 TeV.
The number of entries per bin where divided by the bin widths.
The distributions have  smoothly falling shapes, similar to those for 2-body invariant masses studied by  ATLAS~\cite{highmass,Aaboud:2017yvp} and CMS~\cite{2017520,Khachatryan:2016ecr,Sirunyan:2018pas}.
Therefore, the $\mjjl$ distributions can be 
examined for local excesses above a data-derived estimate of the smoothly falling predictions, which can be obtained
using various smoothing techniques or performing a global fit with an analytic function.

\begin{figure}[h]
\begin{center}
   \includegraphics[width=0.70\textwidth]{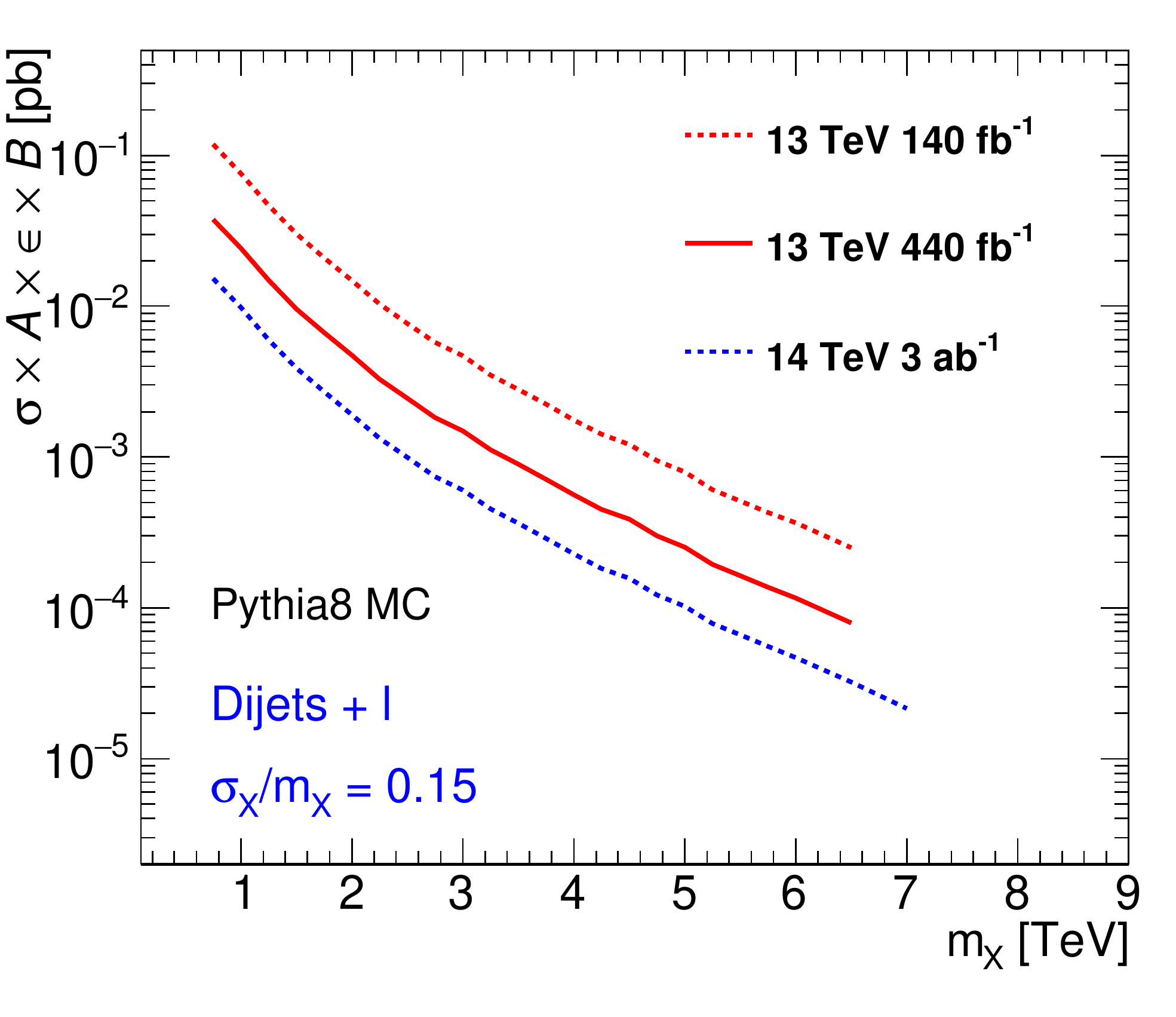}
\end{center}
\caption{
The 95\% CL upper limits obtained from the $\mjjl$ distribution on cross-section times acceptance ($A$), efficiency ($\epsilon$) and branching ratio (BR), for a BSM signal with a cross-section that produces a Gaussian contribution to the particle-level $\mjjl$ distribution, as a function of the mean of the Gaussian mass distribution. The limits are calculated for a BSM signal assuming $\sigma_X / m_X=0.15$ width.
}
\label{fig:LHCmjjlLimits}
\end{figure}

Figure~\ref{fig:LHCmjjlLimits} shows the expected $95\%$ confidence-level (C.L.) upper limit
on fiducial cross-section times the  branching ratio for a generic Gaussian signal with the width ($\sigma_G$) being  15\% of the Gaussian peak position, while decaying to two jets and a lepton. 
The results are presented for different values of the integrated luminosity, including luminosity expected after the LHC run 3 and high-luminosity LHC.
The limits are calculated using the frequentist approach \cite{Cowan:2010js} with the asymptotic approximation \footnote{Due to the approximate nature of this study, we did not take into account systematic effects in the limit calculation. For 2-jet plus lepton (or two lepton) invariant masses, systematic uncertainties for exclusion limits are expected to be dominated by jet energy scale and jet energy resolution. Typically, such experimental effects are not significant (within the 1-sigma band on the limits \cite{Aad:2020kep}).}.
The $95\%$ quantile of the posterior is taken as the upper limit on the possible
number of signal events in data corresponding to that mass point.
This value, divided by the corresponding luminosity, provides
the upper limit on the production cross section of a new particle times the branching ratio (Br) to two jets plus lepton.
The presented limits can be used to estimate the sensitivity of the LHC
data to any phenomenological models predicting Gaussian signals in 
the $\mjjl$ distributions.

As emphasized above, the limits for the 4-body masses, such as $\mjjll$ or $\mjjjl$, are difficult to obtain due to the lack of experimental publications that can be used
to correct Monte Carlo predictions for  realistic lepton misidentification rates.

\section{Summary}

This paper discusses several scenarios for model-independent BSM searches that have not received    
a sufficient attention in  collider experiments. 
In particular, we focus on generic processes  with the cascade decays of the type 
$A \rightarrow B C \rightarrow jjl(l)$, where $A$ is a heavy 
BSM particle, while either $B$ or $C$ is another heavy particle decaying to jets with a mass smaller than $A$.
For the physics processes discussed in this paper, multi-body 
invariant masses calculated using two jets and leptons may lead  to favourable conditions for experimental 
observations of signal-like structures on a smoothly falling  invariant masses.
Such structures can be observed with high precision using various soothing numerical techniques or 
a global fit of multi-body invariant masses
using analytic functions. All such methods are widely utilized in dijet searches.

We believe that multi-body final states with leptons are largely unexplored using 
high-precision model-independent techniques that do not  rely on  Monte Carlo predictions for SM background.
It would be interesting to explore general theoretical scenarios with narrow heavy particles that
decay to relatively broad resonances.
Such BSM events can easily escape observations using previously published searches
in dijets, which are not the primary observables for such BSM events. Dijets searches in inclusive collision events are also affected by the overwhelming multijet QCD background.
Therefore, multi-body invariant masses that include lepton 
have the potential to bring unexpected discoveries. Using Monte Carlo simulations,
our analysis  sets 95\% credibility-level
upper limits on the signal cross-section times acceptance times efficiency times
branching ratio for new processes that can produce a Gaussian contribution to
the dijet+lepton invariant-mass distribution. The estimated limits 
can be used for guiding model builders who hypothesize heavy states leading 
to signal signatures in such multi-body invariant masses.

The approach described in this paper can equally be applied to the invariant masses that include more than four objects, or invariant masses that are composed from only jets or leptons. However, such studies are beyond the scope of the current paper.

\section*{Acknowledgments}
We would like to thank K.~S.~Agashe, H.~Meng, P.~Du and M.~Chala for the discussions of Monte Carlo simulations for the radion and composite BSM models. W.Islam would like to thank A.~Khanov also for some useful discussions. We gratefully acknowledge the computing resources provided by
the Laboratory Computing Resource Center at Argonne National Laboratory.
The submitted manuscript has been created by UChicago Argonne, LLC, Operator of Argonne National Laboratory (“Argonne”). Argonne, a U.S. 
Department of Energy Office of Science laboratory, is operated under Contract No. DE-AC02-06CH11357. The U.S. Government retains for itself, 
and others acting on its behalf, a paid-up nonexclusive, irrevocable worldwide license in said article to reproduce, prepare derivative works, 
distribute copies to the public, and perform publicly and display publicly, by or on behalf of the Government.
The Department of Energy will provide public access to these results of federally sponsored research in accordance with the 
DOE Public Access Plan. \url{http://energy.gov/downloads/doe-public-access-plan}. Argonne National Laboratory’s work was 
funded by the U.S. Department of Energy, Office of High Energy Physics under contract DE-AC02-06CH11357. 


%



\begin{thebibliography}{10}




\bibitem[Author(year)]{highmass} ATLAS Collaboration. Search for new phenomena in dijet mass and angular distributions from $pp$ collisions at $\sqrt{s}=$ 13 TeV with the ATLAS detector. {\emph{Phys. Lett.}} {\bf 2016}, {\em b754}, 302 -- 322  
doi:10.1016/j.physletb.2016.01.032.

 



\bibitem[Author(year)]{Aaboud:2017yvp} ATLAS Collaboration. Search for new phenomena in dijet events using 37 $fb^{-1}$ of $pp$ collision data collected at $\sqrt{s}=$13 TeV with the ATLAS detector. {\emph{Phys. Rev.}} {\bf 2017}, {\em D96}, 052004 doi:10.1103/PhysRevD.96.052004.



\bibitem[Author(year)]{2017520} CMS Collaboration. Search for dijet resonances in proton–proton collisions at 13 TeV and constraints on dark matter and other models. {\emph{Phys. Lett.}} {\bf 2017}, {\em b769}, 520-542  doi:10.1016/j.physletb.2017.09.029.


\bibitem[Author(year)]{Khachatryan:2016ecr} CMS Collaboration. Search for narrow resonances in dijet final states at $\sqrt(s)=$ 8 TeV with the novel CMS technique of data scouting. {\emph{Phys. Rev. Lett.}} {\bf 2016}, {\em 117}, 031802 doi:10.1103/PhysRevLett.117.031802.


\bibitem[Author(year)]{Sirunyan:2018pas} CMS Collaboration. Search for narrow resonances in the b-tagged dijet mass spectrum in proton-proton collisions at $\sqrt{s} =$ 8 TeV. {\emph{Phys. Rev. Lett.}} {\bf 2018}, {\em 120}, 201801 doi:10.1103/PhysRevLett.120.201801.



\bibitem[Author(year)]{Chatrchyan:2012uxa} CMS Collaboration. Search for Three-Jet Resonances in $pp$ Collisions at $\sqrt{s}=7$ TeV. {\emph{Phys. Lett.}} {\bf 2012}, {\em 718}, 329--347 doi:10.1016/j.physletb.2012.10.048.


\bibitem[Author(year)]{Sirunyan:2018duw} CMS Collaboration. Search for pair-produced three-jet resonances in proton-proton collisions at $\sqrt s$ =13  TeV. {\emph{Phys. Lett.}} {\bf 2019}, {\em 99}, 012010 doi:10.1103/PhysRevD.99.012010.


\bibitem[Author(year)]{Accomando:2019ahs} Accomando, E.; Coradeschi, F.; Cridge, T.;  Fiaschi, J.; Hautmann, F.; Moretti, S.; Shepherd-Themistocleous, C.; Voisey, C. Production of Z'-boson resonances with large width at the LHC. {\emph{Phys. Lett.}} {\bf 2020}, {\em 803}, 135293 doi:10.1016/j.physletb.2020.135293.


\bibitem[Author(year)]{Li:2019pag} Li, Y.Y.; Nicolaidou, R.; Paganis, S. Exclusion of heavy, broad resonances from precise measurements of $WZ$ and $VH$ final states at the LHC. {\emph{Eur. Phys. J.}} {\bf 2019}, {\em 79}, 348 doi:10.1140/epjc/s10052-019-6858-5.


\bibitem[Author(year)]{Aaboud:2019zxd} ATLAS Collaboration. Search for low-mass resonances decaying into two jets and produced in association with a photon using $pp$ collisions at $\sqrt{s} = 13$ TeV with the ATLAS detector. {\emph{Phys. Lett.}} {\bf 2019}, {\em 795}, 56 -- 75 doi:10.1016/j.physletb.2019.03.067.


\bibitem[Author(year)]{Sirunyan:2019sgo} CMS Collaboration. Search for Low-Mass Quark-Antiquark Resonances Produced in Association with a Photon at $\sqrt {s}$ =13  TeV. {\emph{Phys. Rev. Lett.}} {\bf 2019}, {\em 123}, 231803 doi:10.1103/PhysRevLett.123.231803.



\bibitem[Author(year)]{Sirunyan:2017nvi} Sirunyan, A.M.; Tumasyan, A.; Adam, W.; Ambrogi, F.; Asilar, E.; Bergauer, T.; Brstetter, J.; Brondolin, E.; Dragicevic, M.; Erö, J.; et~al. Search for low mass vector resonances decaying into quark-antiquark pairs in proton-proton collisions at $ \sqrt{s}=13 $ TeV. {\emph{JHEP}} {\bf 2018}, {\em 1}, 1--41, doi:10.1007/JHEP01(2018)097.


\bibitem[Author(year)]{Sirunyan:2017dnz} CMS Collaboration. Search for Low Mass Vector Resonances Decaying to Quark-Antiquark Pairs in Proton-Proton Collisions at  13 TeV. {\emph{Phys. Rev. Lett.}} {\bf 2017}, {\em 119}, 111802 doi:10.1103/PhysRevLett.119.111802.



\bibitem[Author(year)]{Sirunyan:2018ikr} CMS Collaboration. Search for low-mass resonances decaying into bottom quark-antiquark pairs in proton-proton collisions at $\sqrt{s} =$ 13 TeV. {\emph{Phys. Rev.}} {\bf 2019}, {\em 99}, 012005  doi:10.1103/PhysRevD.99.012005.



\bibitem[Author(year)]{Sirunyan:2019vxa} CMS Collaboration. Search for low mass vector resonances decaying into quark-antiquark pairs in proton-proton collisions at $\sqrt{s}=$ 13 TeV. {\emph{Phys. Rev.}} {\bf 2019}, {\em 100}, 1--41, doi:10.1103/PhysRevD.100.112007.


\bibitem[Author(year)]{Dorigo:2018cbl} Dorigo, T. Hadron Collider Searches for Diboson Resonances. {\emph{Prog. Part. Nucl. Phys.}} {\bf 2018}, {\em 100}, 211 -- 261 doi:10.1016/j.ppnp.2018.01.009.



\bibitem[Author(year)]{Aad:2020kep} ATLAS Collaboration. Search for dijet resonances in events with an isolated charged lepton using $\sqrt{s} = 13$ TeV proton-proton collision data collected by the ATLAS detector. {\emph{JHEP}} {\bf 2020}, {\em 6}, 151, doi:10.1007/JHEP06(2020)151.


\bibitem[Author(year)]{Sjostrand:2006za} Sjöstr, T.; Mrenna, S.; Skands, P.  PYTHIA 6.4 Physics and Manual. {\emph{JHEP}} {\bf 2006}, {\em 5}, 26  




\bibitem[Author(year)]{Sjostrand:2007gs} Sjöstrand, T.; Mrenna, S.; Skands, P.  A Brief Introduction to PYTHIA 8.1. {\emph{Comput. Phys. Commun.}} {\bf 2008}, {\em 178}, 852--867, doi:10.1016/j.cpc.2008.01.036.



\bibitem[Author(year)]{Ball:2014uwa} Ball, R.D.; Bertone, V.; Carrazza, S.; Deans, C.S.; Del Debbio, L.; Forte, S.; Guffanti, A.; Hartl, N.P.; Latorre, J.I.; Rojo, J.; Ubiali, M. Parton distributions for the LHC Run II. {\emph{JHEP}} {\bf 2015}, {\em 4}, 40, doi:10.1007/JHEP04(2015)040.


\bibitem[Author(year)]{Buckley:2014ana} Buckley, A.; Ferr, o J.; Lloyd, S.; Nordström, K.; Page, B.; Rüfenacht, M.; Schönherr, M.; Watt, G.  LHAPDF6: Parton density access in the LHC precision era. {\emph{Eur. Phys. J.}} {\bf 2015}, {\em 75}, 1--20, doi:10.1140/epjc/s10052-015-3318-8.



\bibitem[Author(year)]{Cacciari:2008gp} Cacciari, M.; Salam, G.P.; Soyez, G. The anti-$k_t$ jet clustering algorithm. {\emph{JHEP}} {\bf 2008}, {\em 4}, 63, doi:10.1088/1126-6708/2008/04/063.


\bibitem[Author(year)]{Cacciari:2011ma} Matteo, C.; Gavin, S.P.; Gregory, S.  FastJet User Manual. {\emph{Eur. Phys. J.}} {\bf 2012}, {\em C72}, 1896, doi:10.1140/epjc/s10052-012-1896-2.


\bibitem[Author(year)]{Chekanov_2018} Chekanov, S.V.; Childers, J.T.; Proudfoot, J.; Wang, R.; 
 Frizzell, D. Precision searches in dijets at the HL-LHC and HE-LHC. {\emph{J. Instrum.}} {\bf 2018}, {\em13}, P05022, doi:10.1088/1748-0221/13/05/p05022.

\bibitem[Author(year)]{Aad:2013gma} ATLAS Collaboration. Search for Quantum Black Hole Production in High-Invariant-Mass Lepton$+$Jet Final States Using $pp$ Collisions at $\sqrt{s} =$ 8  TeV and the ATLAS Detector. {\emph{Phys. Rev. Lett.}} {\bf 2014}, {\em 112}, 091804. doi:10.1103/PhysRevLett.112.091804.


\bibitem[Author(year)]{Agashe:2016kfr} Agashe, K.S.; Collins, J.H.; Du, P.; Hong, S.; Kim, D.; Mishra, R.K. LHC Signals from Cascade Decays of Warped Vector Resonances. {\emph{JHEP}} {\bf 2017}, {\em 5}, 78, doi:10.1007/JHEP05(2017)078.


\bibitem[Author(year)]{Chala_2018} Mikael, C.; Michael, S. Behavior of composite resonances breaking lepton flavor universality. {\emph{Phys. Rev.}} {\bf 2018}, {\em 98}, 035010, doi:10.1103/physrevd.98.035010.

\bibitem[Author(year)]{Alwall_2014} Alwall, J.; Frederix, R.; Frixione, S.; Hirschi, V.; Maltoni, F.; Mattelaer, O.; Shao, H.-S.; Stelzer, T.; Torrielli, P.; Zaro, M. The automated computation of tree-level and next-to-leading order differential cross sections, and their matching to parton shower simulations. {\emph{JHEP}} {\bf 2014}, {\em 7}, 79, doi:10.1007/JHEP07(2014)079.


\bibitem[Author(year)]{Chekanov:2017pnx} Chekanov, S. V., HepSim: a repository with predictions for high-energy physics experiments.
{\emph{Advances in High Energy Physics}},  {\bf 2015}, {\em 2015},  136093, doi:10.1155/2015/136093.   

\bibitem[Author(year)]{Cowan:2010js} Cowan, G.; Cranmer, K.; Gross, E.; Vitells, O. Asymptotic formulae for likelihood-based tests of new physics. {\emph{Eur. Phys. J.}} {\bf 2011}, {\em 71}, 1554, doi:10.1140/epjc/s10052-011-1554-0.


\end{thebibliography}
\end{document}